\documentclass[fleqn,usenatbib]{mnras}
\usepackage{newtxtext,newtxmath}
\usepackage[T1]{fontenc}

\DeclareRobustCommand{\VAN}[3]{#2}
\let\VANthebibliography\thebibliography
\def\thebibliography{\DeclareRobustCommand{\VAN}[3]{##3}\VANthebibliography}

\usepackage{graphicx}   % Including figure files
\usepackage{amsmath}    % Advanced maths commands

\usepackage{amssymb}    % Extra maths symbols
\usepackage{pdflscape}
\usepackage{multirow}
\usepackage{threeparttable}
\title[Distribution of $GRBs$ on the $T_{90}-HR$ plane]{Distribution of Gamma-Ray Bursts on the $\mathbf{T_{90}-HR}$ plane and Their Classification Revisited}

\author[L. Zhang et al.]{
Liang Zhang$^{1}$\thanks{E-mail: liang\_zhang\_gz@sina.com},
Juan-Juan Luo$^{2}$\thanks{E-mail: j\_j\_luo@sina.com},
Yong-Feng Huang$^{3,4}$\thanks{E-mail: hyf@nju.edu.cn},
Yu-Jun Gong$^{5}$,
and Sheng Wu$^{6}$
\\
$^{1}$Guizhou Vocational College of Economics and Business, Duyun 558022 , P. R. China \\
$^{2}$School of Physics and Electronics, Qiannan Normal University for Nationalities, Duyun 558000, P. R. China\\
$^{3}$School of Astronomy and Space Science, Nanjing University, Nanjing 210023, P. R. China \\
$^{4}$Key Laboratory of Modern Astronomy and Astrophysics (Nanjing University), Ministry of Education, Nanjing 210023, P. R. China \\
$^{5}$Qiannan Polytechnic for Nationalities, Duyun 558022, P. R. China \\
$^{6}$Guangzhou Intelligence Communications Technology Co., Ltd, Guangzhou 510630, P. R. China
}

\date{Accepted XXX. Received YYY; in original form ZZZ}

\pubyear{2015}

\begin{document}
\label{firstpage}
\pagerange{\pageref{firstpage}--\pageref{lastpage}}
\maketitle
\begin{abstract}
Using four mixed bivariate distributions ($Normal$ distribution,
$Skew$-$Normal$ distribution, $Student$ distribution,
$Skew$-$Student$ distribution) and bootstrap re-sampling analysis,
we analyze the samples of $CGRO/BATSE$, $Swift/BAT$ and
$Fermi/GBM$ gamma-ray bursts in detail on the $T_{90} - HR$
(Hardness Ratio) plane. The Bayesian information criterion is used
to judge the goodness of fit for each sample, comprehensively. It
is found that all the three samples show a symmetric (either
$Normal$ or $Student$) distribution. It is also found that the
existence of three classes of gamma-ray bursts is preferred by the
three samples, but the strength of this preference varies with the
sample size: when the sample size of the data set is larger, the
preference of three classes scheme becomes weaker. Therefore, the
appearance of an intermediate class may be caused by a small
sample size and the possibility that there are only two classes of
gamma-ray bursts still cannot be expelled yet. A further bootstrap
re-sampling analysis also confirms this result.

\end{abstract}% Select between one and six entries from the list of approved keywords.
%%%%%%%%%%%%%%%%% BODY OF PAPER %%%%%%%%%%%%%%%%%%

\section{INTRODUCTION}

\label{sec:INTRODUCTION} Gamma-ray bursts (GRBs;
\citealt{Klebesadel1973ApJ...182L..85K}) are the most violent
stellar explosions in the universe. The duration of $T_{90}$ is
defined as the time corresponding to 5\% -- 95\% of a burst
fluence \citep{1993ApJ...413L.101K,1995AAS...186.5301K,2016LPICo1962.4099H}.
Largely based on the parameter of $T_{90}$, the short and long GRBs
classification sketch was established \citep{1989Natur.340..126E}
and intensive researches
\citep{1993ApJ...413L.101K,1998ApJ...494L..45P,2007MNRAS.374L..34K,2012ApJ...756...44R,
2012wds..conf..129B,2015A&A...581A..29T,2016NewA...46...54T,2016Ap&SS.361..125T,
2016MNRAS.458.2024T,2015Ap&SS.357....7Z,2018Ap&SS.363..223Z,2016PASJ...68S..30O,
2016MNRAS.462.3243Z,2017Ap&SS.362...70K,2018MNRAS.473..625K} have
shown that $T_{90} \approx 2s$ is the critical duration of this
classification. For the origin of $GRBs$, the mainstream view is
that long Gamma-Ray Bursts (lGRBs) with a duration of $T_{90}
> 2s$ originate from the collapes of massive stars
\citep{1999Natur.401..453B,2003ApJ...599..394M,1993ApJ...405..273W,1998ApJ...494L..45P,
2000ApJ...537..810W}, while short Gamma-Ray Bursts (sGRBs) with
$T_{90} < 2s$ come from double neutron star (NS-NS) or NS-black
hole (BH) mergers
\citep{1989Natur.340..126E,2007PhR...442..166N,Tanvir2013Natur.500..547T,
Goldstein2017ApJ...848L..14G,Abbott2017ApJ...848L..12A}. However,
note that the possibility that some GRBs may be produced by other
processes than the above two mechanisms still cannot be completely
expelled yet. For example, some events may be associated with the
kick of high speed neutron stars
\citep{2003ApJ...594..919H,2022MNRAS.509.4916X}. Consequently,
there may also exist other kinds of GRBs. Examining the
classification of GRBs can help reveal their trigger mechanisms.

The $T_{90}$ duration was noticed early on to be composed of two
lognormal functions
\citep{1994MNRAS.271..662M,1996ApJ...463..570K,1996AIPC..384...42K}.
However, \citet{1998ApJ...508..757H} found a prominent third group
between the classic short and long groups when studying the
$T_{90}$ data of 797 $CGRO/BATSE$ GRBs, which means the emergence
of the intermediate-duration class of GRBs. The third class was
reconfirmed by different data sets from various detectors such as
$Swift/BAT$ \citep{2008A&A...489L...1H} and $Fermi/GBM$
\citep{2015Ap&SS.359...20T}. Several authors
\citep{2009BaltA..18..311H,2015Ap&SS.357....7Z,2016MNRAS.462.3243Z}
analyzed those GRBs with measured redshifts in both the observer
frame and the rest frame. They argued that the three Gaussian
component model and the two Gaussian component model are almost
equally suitable. \citet{2011A&A...525A.109D} carefully analyzed
the properties of the presumed intermediate GRBs and found that
they differ from long GRBs only in having a lower luminosity, so
that they might be simply a low-luminosity tail of the lGRB group
\citep{2011ApJ...739L..55B}. Also note that this intermediate GRBs
may relate to short GRBs with extended emission (sGRBEE;
\citealt{2006ApJ...643..266N,2021ApJ...911L..28D}), i.e., they
have a duration that would make them being identified as long GRBs
but without an associated supernova. They could most likely
originate from the merger of a white dwarf with an NS
\citep{2007MNRAS.374L..34K} or BH \citep{2018MNRAS.475L.101D}.

Further separation of long GRBs into subgroups was also
considered. \citet{2002A&A...392..791H} studied a large sample of
2041 $CGRO/BATSE$ GRBs and found that the elusive third group
seems to be blended into the long GRB group. A similar conclusion
is also drawn by \citet{2015Ap&SS.359...20T}. It is worth noting
that \citet{2015A&A...581A..29T,2019ApJ...870..105T} attributed
this phenomenon to the skewness of the component.
\citet{2015A&A...581A..29T} claimed that the logarithmic duration
distribution need not necessarily be symmetrical. The asymmetry
(skewness) can originate from, e.g., an asymmetric distribution of
the progenitor envelope mass \citep{2015Ap&SS.357....7Z}. Several
authors
\citep{2016Ap&SS.361..125T,2016MNRAS.458.2024T,2018MNRAS.473..625K}
have tested the skew distribution of $BATSE$, $Swift$ and $Fermi$
GRBs separately. It was found that the mixture of two skewed
components are either significantly better than, or at least as
good as, three-component symmetric models, indicating that the
third class is unnecessary and could be discarded.

It has always been a controversial topic that GRBs can be
classified into two components or else, by univariate analysis. A
natural idea is to introduce bivariate analysis. Many studies have
introduced the ratio of fluence in different bands, also known as
hardness ratio (hereafter $HR$), as the second variable to study
the classification of GRBs on the $log T_{90} - log HR$ plane.
However, different results are obtained by different groups.
\citet{1998ApJ...508..314M} and
\citet{Horv2006A&A...447...23H,2010ApJ...713..552H} argued that
three Gaussian components are the optimal interpretation for GRBs
on the $log T_{90} - log HR$ plane, while
\citet{2012ApJ...756...44R}, \citet{2016MNRAS.462.3243Z} and
\citet{2016ApJS..223...28N} suggested that two components are
favored by the observational data. Very recently,
\citet{2019MmSAI..90...45T,2019ApJ...870..105T} used four kinds of
bivariate distribution functions -- normal, skew-normal, Student,
skew-Student (hereafter $\mathcal{N}$, $\mathcal{S N}$,
$\mathcal{T}$, $\mathcal{S T}$, respectively) to fit the data sets
of $CGRO/BATSE$, $Fermi/GBM$, $Swift/BAT$, $Konus-Wind$, $RHESSI$,
$Suzaku/WAM$ GRBs on the $logT_{90} - log HR$ plane. The presence
of two or three components varies from detector to detector. It
should be emphasized that the motivation of introducing skewed
distribution is due to the fact that the original data set itself
might be skewed \citep{2019MmSAI..90...45T,2019ApJ...870..105T}.
In this case, modeling an inherently skewed distribution with a
mixture of symmetric groups requires excessive components to be
included, resulting in a spurious determination of the number of
underlying classes \citep{2012MNRAS.420..405K}.

In previous studies, when skewed distribution functions are used
to fit the observational data of different missions, the sample
sizes are usually very limited,  especially for $Fermi/GBM$ and
$Swift/BAT$ GRBs \citep{2019MmSAI..90...45T,2019ApJ...870..105T}.
Here we follow \citet{2019MmSAI..90...45T,2019ApJ...870..105T} to
study the distribution and classification of GRBs on the $T_{90} -
HR$ plane for $CGRO/BATSE$, $Swift/BAT$ and $Fermi/GBM$ events,
with the sample sizes significantly expanded. To increase the
stability and credibility of the analysis, we also adopt the
statistical bootstrap re-sampling method as done by
\citet{2016MNRAS.462.3243Z}.

Our paper is organized as follows. The data selection process is
described in Section 2. Our statistical methods and the four
bivariate distribution functions used in this study are introduced
in Section 3. The main results are presented in Section
\ref{sec:RESULT}.
%% Section \ref{sec:DISCUSSION} is a brief discussion on the classification of GRBs.
Finally, Section \ref{sec:CONCLUSION} is our conclusion and
discussion. \textbf{R}\footnote{https://cran.r-project.org/}
statistical language is utilized throughout the paper and the
fittings of observational data points are performed with the
\textbf{R} package of
\textbf{mixsmsn}\footnote{https://cran.r-project.org/web/packages/mixsmsn/index.html}
\citep{JSSv054i12}.

\section{DATA SELECTION}
\label{sec:DATA SELECTION AND METHOD}

We use three GRB samples in our study. These events are detected
by $CGRO/BATSE$, $Swift$, and $Fermi$, respectively. Here we
describe the three data sets as follows.

(i) The $CGRO/BATSE$
catalogue\footnote{https://heasarc.gsfc.nasa.gov/W3Browse/cgro/batsegrb.html}
contains a total of 2702 GRBs. Among these events, 1954 events
have valid $T_{90}$, $Fluence_2 (50-100 keV)$, $Fluence_3 (100-300
keV)$ data, which are selected for our study. The $HR$ is
calculated as

\begin{equation}\label{eq:1}
HR=\frac{S_{100-300keV}}{S_{50-100keV}}=\frac{{Fluence_3}}{Fluence_2}.
\end{equation}

(ii) The $Swift$ satellite, dedicated to GRBs studies, was
launched on 2004 November 20
(\citealt{Gehrels2004,Butler2007,2021yCat..18969020J}). The
$Swift/BAT$ catalogue contains 1526 GRBs as of March 28, 2022
\footnote{https://swift.gsfc.nasa.gov/archive/grb\_table/}.
However, it does not provide fluence values
in the required energy bands needed to calculate the $HR$
parameter. In fact $Swift$ catalog contains fluence values,
but only in one band from 15 to 150 keV. So, we use the optimal
spectral fitting model to derive $HR$. When the optimal
fitting model is a cutoff power-law (CPL) function, the
corresponding $Epeak$ value is then obtained from the GCN
circular\footnote{https://gcn.gsfc.nasa.gov/gcn3\_archive.html}
\footnote{https://www.mpe.mpg.de/$\sim$jcg/grbgen.html}, such as
in the cases of GRBs 220325A, 220101A, 211225b, etc. Finally, we
obtained a sample of 1365 $Swift/BAT$ GRBs with necessary
parameters available. The parameter of $HR$ is then calculated as

\begin{equation}\label{eq:2}
HR=\frac{S_{50-100keV}}{S_{15-25keV}}=\frac{\int^{100keV}_{50keV}F(E)EdE}{\int^{25keV}_{15keV}F(E)EdE},
\end{equation}
where $f(E)$ corresponds to the best fit spectrum function which
could be a power-law (PL) or CPL. % or a Comptonized (COMP) model.

(iii) The $Fermi$ satellite, dedicated to high energy phenomena
and GRB studies, was launched in June, 2008 (
\citealt{Meegan2009}). The $Fermi/GBM$ catalogue
\citep{2020ApJ...893...46V} contains 3255 GRBs as of March 28,
2022, which can be accessed through the HEASARC
website\footnote{https://heasarc.gsfc.nasa.gov/W3Browse/fermi/fermigbrst.html}.
Similar to the $Swift$ GRB catalogue, the $Fermi$ catalogue also
does not provide required fluence data that could be directly
used to calculate the hardness parameter. Again, we resort to the
optimal spectral fitting model. The optimal spectral fittings are
available for 2310 $Fermi$ GRBs, of which the spectra are best fit
by one of the four forms: power-law ('plaw'), Comptonized
('comp'), Band ('band')
\citep{1993AAS...182.7409B,1993ApJ...413..281B}, and smoothly
broken power law ('sbpl') \citep{1999ApL&C..39..281R} . Using the
spectrum, the $HR$ parameter is then calculated as:

\begin{equation}\label{eq:3}
HR = \frac{S_{100-300keV}}{S_{25-50keV}} =
\frac{\int^{300keV}_{100keV}F(E)EdE}{\int^{50keV}_{25keV}F(E)EdE}.
\end{equation}

For the above three GRB samples, we have downloaded the duration
data from the corresponding websites and calculated the $HR$
parameter.   Table \ref{tab:table1} sums up the general features
of our three data sets.

\begin{table*}\normalsize
\centering \caption{Basic features of the three GRB samples used
in this study }
\label{tab:table1}
\begin{threeparttable}
\begin{tabular}{lccclcc}
\hline
Data Set Name & No.\tnote{a} & Source & Model\tnote{b} & Parameters\tnote{c} & Energy range (keV) \tnote{d} & Reference \\
\hline
$CGRO/BATSE$ & 1954 (2702) & $BATSE$ catalogue & - & $T_{90}$, fluence2, fluence3 & 100 --- 300 , 50 --- 100 & e\\

\multirow{2}{*}{$Swift/BAT$} & \multirow{2}{*}{1365 (1526)} & \multirow{2}{*}{$Swift$ catalogue} & PL & $T_{90}$, index & \multirow{2}{*}{50 --- 100 , 15 --- 25}  & \multirow{2}{*}{f, g}\\
 & & & CPL & $T_{90}$, index, Epeak & & \\

\multirow{4}{*}{$Fermi/GBM$} & \multirow{4}{*}{2310 (3255)} & \multirow{4}{*}{$Fermi$ catalogue} & plaw & $T_{90}$, index & \multirow{4}{*}{300 --- 100 , 50 --- 25}  & \multirow{4}{*}{h}\\
& & & comp & $T_{90}$, index, Epeak & & \\
& & & band & $T_{90}$, alpha, beta, Epeak & & \\
& & & sbpl & $T_{90}$, index1, brken, brksc, index2 & & \\
\hline
\end{tabular}
\begin{tablenotes}
       \footnotesize
       %\item[] Notes.
       \item[a] The number of GRBs used in this study
                (i.e. the number of GRBs having valid $T_{90}$ and $HR$ parameters.
                The number in the parentheses represents the total number of GRBs detected
                by each detector as of March 28, 2022.
       \item[b] The optimal spectral fitting models available on the website. We use the
                optimal spectral fitting model to calculate the $HR$ parameter when the required
                fluences are not directly available.
       \item[c] Parameters involved in the optimal spectral fitting model.
       \item[d] Energy ranges defined for calculating the $HR$ parameter.
       \item[e] website: https://heasarc.gsfc.nasa.gov/W3Browse/cgro/batsegrb.html
       \item[f] website: https://swift.gsfc.nasa.gov/archive/grb\_table/
       \item[g] website: https://gcn.gsfc.nasa.gov/gcn3\_archive.html
       \item[h] website: https://heasarc.gsfc.nasa.gov/W3Browse/fermi/fermigbrst.html
     \end{tablenotes}
\end{threeparttable}
\end{table*}

\section{Method}
\label{sec:Method}

\subsection{Maximum likelihood Algorithm}
\label{sec:Maximum Loglikelihood Algorithm}

For a given distribution, we denote the probability density
function (hereafter PDF) as $f (\boldsymbol{x}; \theta)$, where
$\theta = \left\{\theta_i\right\} ^p_{i=1}$ is a set of parameters
\citep{2019ApJ...870..105T}. The logarithmic likelihood function
is:

\begin{equation}\label{eq:4}
\mathcal{L}_{p}(\theta)= \sum_{i=1}^{N} \ln f\left(x_i ; \theta \right),
\end{equation}
where $\left\{\boldsymbol{x}_i\right\}^N_{i=1}$ is the data sample
based on which a distribution function is tested
\citep{2019ApJ...870..105T}. The maximum likelihood (hereafter ML)
algorithm \citep{2022MNRAS.513L...1B,2022icrc.confE.768C} tries to
find the largest probability value. Therefore, the goal is to find
a set of $\hat{\theta}$ to maximize the likelihood function, i.e.,
$\mathcal{L}_{p,max} \equiv \mathcal{L}_{p}(\hat{\theta})$.

\subsection{Four Mixed Bivariate Distributions}

\label{sec:Four Mixed Bivariate Distributions}

The total PDF of a mixture of $n$ components, each having a PDF
given by $f_{i}(\boldsymbol{x}; \theta^{(i)})$, is defined as

\begin{equation}\label{eq:5}
f(\boldsymbol{x}; \theta) = \sum_{i=1}^{n}A_i
f_i\left(\boldsymbol{x} ; \theta^{(i)}\right),
\end{equation}
where the weights $A_i$ satisfies the relation of $\sum_{i=1}^{n}
A_{i}=1$, $\theta=\bigcup_{i=1}^{n} \theta^{(i)}$
\citep{2019ApJ...870..105T}. In this study, we consider four kinds
of distributions, as described below.

The PDF of an $\mathcal{N}$ distribution
\citep{Mudelsee2014} is:

\begin{equation}\label{eq:6}
f^{\mathcal{N}} \left( x; \mu, \Sigma \right) = \frac{1}{2\pi\sqrt{|\Sigma|}} exp \left[-\frac{1}{2} \left(x - \mu \right)^{T} \Sigma^{-1} \left( x - \mu \right) \right],
\end{equation}
where $\mu$ is the mean of the parameter set,  $| \Sigma | = det(
\Sigma )$, and $\Sigma$ is the covariance matrix,
\citep{2019ApJ...870..105T},

\begin{equation}\label{eq:10}
\Sigma = \left(
\begin{array}{cc}
\sigma_{x}^{2}&\rho\delta_{x}\sigma_{y}\\
\rho\delta_{x}\sigma_{y}&\sigma_{y}^{2}
\end{array}
\right).
\end{equation}
A mixture of $n$ components is described by $p = 6n - 1$ free parameters.

Similarly, the PDF of an $\mathcal{SN}$ distribution \citep{1986A}
is

\begin{equation}\label{eq:7}
f^{ \mathcal{S N} } \left( x ; \mu, \Sigma, \lambda \right) = 2f^{ \mathcal{N} } \left( x ; \mu, \Sigma \right) \Phi \left(\lambda^{T} \Sigma^{-1/2} \left( x -\mu \right) \right),
\end{equation}
where $\Phi$ represents the cumulative distribution function (CDF) of a univariate standard normal
distribution, $\boldsymbol{\lambda}$ denotes the skewness
parameter \citep{2019ApJ...870..105T}. A mixture of $n$ components
is described by $p = 8n - 1$ free parameters.

The PDF of a $\mathcal{T}$ distribution
\citep{doi:10.1080/01621459.1989.10478852} is

\begin{equation}\label{eq:8}
\begin{aligned}
f^{ \mathcal{T} }(x ; \mu, \Sigma, \nu)=& \frac{1}{\pi \nu \sqrt{|\Sigma|}} \frac{\Gamma\left(\frac{\nu+2}{2}\right)}{\Gamma\left(\frac{\nu}{2}\right)} \\
& \times\left(1+\frac{1}{\nu}(x-\mu)^{\top} \Sigma^{-1}(x-\mu)\right)^{-\frac{\nu+2}{2}},
\end{aligned}
\end{equation}
where $\nu$ is degrees of freedom (dof) and $\Gamma$ is the gamma
function.  $\mu$ is the mean of the $\mathcal{T}$ distribution,
and the covariance matrix is
$\frac{\nu}{\nu-2}\boldsymbol{\Sigma}$
\citep{2019ApJ...870..105T}. A mixture of $n$ components is
described by $p = 6n$ free parameters.

The PDF of an $\mathcal{S T}$ distribution is
\citep{https://doi.org/10.1111/1467-9868.00391}

\begin{equation}\label{eq:9}
\begin{aligned}
&f^{\mathcal{S T}}(x ; \mu, \Sigma, \nu, \lambda )=2 f^{(\mathcal{T})}(x ; \mu, \Sigma, \nu) \\
&\quad \times T_{\nu+2}\left(\sqrt{\frac{\nu+2}{\nu+(x-\mu)^{\top} \Sigma^{-1}(x-\mu)}} \lambda^{\top} \Sigma^{-1 / 2}(x-\mu)\right),
\end{aligned}
\end{equation}
where $T_{\nu+2}$ represents the CDF of the standard bivariate
student distribution, $\lambda$ is the skewness parameter vector.
A mixture of $n$ components is described by $p = 8n$ free
parameters.

\begin{table}\normalsize
\centering \caption{The number of free parameters in the four
distributions. } \label{tab:table2}
\begin{threeparttable}
\begin{tabular}{ccl}
\hline
Model name &  & No. of free parameters \tnote{a} \\
\hline
$\mathcal{N}$&  & $6n - 1$ \\
$\mathcal{S N}$&  & $8n - 1$ \\
$\mathcal{T}$&  & $6n$ \\
$\mathcal{S T}$&  & $8n$ \\
\hline
\end{tabular}
\begin{tablenotes}
       \footnotesize
       \item[a] $n$ represents the number of mixed components.
     \end{tablenotes}
\end{threeparttable}
\end{table}

Table \ref{tab:table2} presents the number of free parameters for
the above four distributions, from which we could clearly see the
complexity of each distribution.

\subsection{The Criteria for Model Selection }

\label{sec:Model Evaluate Criteria}

As is mentioned in Section \ref{sec:Maximum Loglikelihood Algorithm} , using the ML algorithm, one can obtain the
best PDF for the observational data by assuming a mixture of $n$
components, with each component follows one of the four kinds of
distributions: $\mathcal{N}$, $\mathcal{S N}$, $\mathcal{T}$,
$\mathcal{S T}$. Theoretically, it is always possible to increase
the logarithmic likelihood value by introducing more free
parameters \citep{2019ApJ...870..105T}, but note that additional
parameters also indicate that the model might be too complicated.
In general, the Akaike information criterion ($AIC$)
\citep{1974ITAC...19..716A} and Bayesian information criterion
($BIC$) \citep{10.1214/aos/1176344136} are widely used as
effective statistical criterions for model selection
\citep{doi:10.1177/0049124104268644,2007JCAP...02..003B,2007MNRAS.377L..74L}.
Herein, the goal is to obtain the simplest possible model that
could adequately describe the data, so we will adopt the $BIC$
criterion. It takes the sample size into consideration to avoid
over-fitting.

The $BIC$ criterion is defined as
\begin{equation}\label{eq:11}
BIC = p \ln N - 2 \ln P_{max},
\end{equation}
where $P_{max}$ is the $ML$ achieved in modeling the data
\citep{2016MNRAS.462.3243Z}. $p$ is the number of free parameters
of the model and $N$ is the sample size.
%From Eq.\ref{eq:11}, it can be seen that for a larger $N$, the $BIC$ criterion prefers those models having a smaller $k$.

The $BIC$ method always chooses the model with the smallest value
$BIC_{min}$ as the best model \citep{2016MNRAS.462.3243Z}. To
assess the goodness of other models, we can calculate their
difference of $BIC$ values with respect to the best model
\citep{doi:10.1177/0049124104268644},
$\Delta{BIC_i}=BIC_i-BIC_{min}$. If $\Delta{BIC_i}<2$, then there
is also substantial support for the $\boldsymbol{i}$th model and
the possibility that it is also a proper description is high. When
$2<\Delta{BIC_i}<6$, then there is significant evidence against
the $\boldsymbol{i}$th model. When $6<\Delta{BIC_i}<10$, the
evidence of rejection is further enhanced. Finally, models with
$\Delta{BIC_i}>10$ yield very strong evidence against the
$\boldsymbol{i}$th model (essentially rejected;
\citep{doi:10.1177/0049124104268644,2007JCAP...02..003B,2016MNRAS.462.3243Z}).

\subsection{Bootstrap Re-sampling Method}

\label{sec:Bootstrap Resampling Method}

The bootstrap re-sampling method \citep{1985daa..conf...81O} is a
powerful statistical tool to quantify the uncertainty associated
with a given estimator or statistical learning method. It
repeatedly draws samples from a training data set and refits a
given model on each sample with the goal of learning more about
the model. \citet{2016MNRAS.462.3243Z} used this methodology to
eliminate the contingency and deviation of their numerical
results. Herein, the bootstrap re-sampling method will also be
used to stabilize the $BIC$ information outputs of the four
models, so as to determine the distribution and classification of
the GRBs observed by the three detectors.

\subsection{Data Analysis Process}

\label{sec:Other Method}

We use the function $\boldsymbol{smsn.mmix()}$ in the
$\boldsymbol{R}$ package $\boldsymbol{mixsmsn}$ to fit the data
sets. This function is responsible for the implementation of the
expectation-maximization (EM) algorithm for the multivariate models, as explained in
Section \ref{sec:Four Mixed Bivariate Distributions}. Note that
the initial value is set by $kmeans$ randomly, so EM algorithm
may converge to different values in each optimization calculation,
resulting in some different $BIC$ values. To overcome this
problem, we use the bootstrap re-sampling method to stabilize the
output $BIC$ values.

\section{Results}
\label{sec:RESULT}

\subsection{Fitting results}
\label{sec:Analysis of fitting results} In this study, the initial
parameters are set as: $\boldsymbol{mu} = NULL$,
$\boldsymbol{Sigma} = NULL$, $\boldsymbol{shape} = NULL$,
$\boldsymbol{pii} = NULL$ and $ get.init = TRUE$. It means that
the initial values of the EM algorithm are obtained using a
combination of the $\boldsymbol{R}$ function $kmeans$ and the
moment method. We further set $error = 0.0001$, which
corresponds to the stopping criterion for the EM algorithm. For
more details, please see \citet{BASSO20102926}.

\begin{table*}
\centering \caption{A direct comparison of $BIC$ between the 1
time fitting results and the $10^4$ times fitting results for the
three GRB samples. } \label{tab:table3}
\begin{threeparttable}
\begin{tabular}{cccccccc}
\hline
\multirow{2}{*}{Detector} & \multirow{2}{*}{Bootstrap time} & \multirow{2}{*}{Component}
& \multirow{2}{*}{$BIC_{min}$\tnote{a}} & \multicolumn{4}{c|}{ $\Delta{BIC}$ ($BIC-BIC_{min}$) of four models  } \\
\cline{5-8}
 & & & & $\mathcal{N}$ & $\mathcal{S N}$ & $\mathcal{T}$ & $\mathcal{S T}$ \\
\hline
\multirow{10}{*}{$CGRO/BATSE$} & \multirow{5}{*}{1} & 1 & \multirow{5}{*}{4848.189} & 756.463 & 354.644 & 727.414 & 299.48 \\
 & & 2 & & 67.624 & 87.654 & 8.622 & 4.592 \\
 & & 3 & & 6.477 & 40.444 & 0 & 45.911 \\
 & & 4 & & 42.670 & 91.662 & 35.746 & 109.456 \\
 & & 5 & & 83.043 & 161.680 & 74.808 & 151.307 \\
 \cline{5-8}
 & \multirow{5}{*}{$10^4$} & 1 & \multirow{5}{*}{4831.242} & 763.845 & 481.017 & 737.251 & 309.055 \\
 & & 2 & & 63.771 & 72.380 & 14.217 & 19.377 \\
 & & 3 & & 1.316 & 38.882 & 0 & 40.518 \\
 & & 4 & & 34.032 & 89.507 & 29.415 & 90.180\\
 & & 5 & & 33.482 & 128.119 & 57.442 & 131.919 \\
\cline{5-8}
\multirow{10}{*}{$Swift/BAT$} & \multirow{5}{*}{1} & 1 & \multirow{5}{*}{2592.555} & 418.816 & 179.351 & 324.3 & 124.69 \\
 & & 2 & & 46.584 & 69.711 & 56.473 & 80.434 \\
 & & 3 & & 0 & 32.961 & 8.865 & 42.884 \\
 & & 4 & & 33.504 & 81.405 & 39.720 & 89.291 \\
 & & 5 & & 62.798 & 130.388 & 70.898 & 137.69 \\
 \cline{5-8}
 & \multirow{5}{*}{$10^4$} & 1 & \multirow{5}{*}{2576.229} & 428.414 & 218.094 & 336.370 & 133.266 \\
  & & 2 & & 52.007 & 71.233 & 63.959 & 79.556 \\
 & & 3 & & 0 & 25.210 & 10.363 & 32.960 \\
 & & 4 & & 25.840 & 65.535 & 34.969 & 73.801 \\
 & & 5 & & 50.677 & 105.976 & 60.654 & 112.524 \\
\cline{5-8}
\multirow{10}{*}{$Fermi/GBM$} & \multirow{5}{*}{1} & 1 & \multirow{5}{*}{5549.238} & 679.57 & 193.943 & 658.199 & 188.664 \\
 & & 2 & & 0 & 38.93 & 5.614 & 43.499 \\
 & & 3 & & 1.968 & 56.071 & 10.755 & 61.587 \\
 & & 4 & & 47.963 & 109.429 & 57.018 & 119.216 \\
 & & 5 & & 85.793 & 153.794 & 95.489 & 160.772 \\
 \cline{5-8}
 & \multirow{5}{*}{$10^4$} & 1 & \multirow{5}{*}{5532.556} & 690.352 & 253.377 & 668.223 & 195.384 \\
 & & 2 & & 5.317 & 40.859 & 10.992 & 42.924 \\
 & & 3 & & 0 & 50.790 & 10.148 & 56.376 \\
 & & 4 & & 32.626 & 90.584 & 42.454 & 89.937 \\
 & & 5 & & 65.993 & 132.932 & 66.188 & 128.966 \\
\hline
\end{tabular}
\begin{tablenotes}
       \footnotesize
       \item[] Note: for the $10^4$ times fitting, $BIC$ is the average value of all the calculations.
       \item[a] The minimum value of $BIC$ for all the calculations.
     \end{tablenotes}
\end{threeparttable}
\end{table*}

For the three GRB samples detected by $CGRO/BATSE$, $Swift/BAT$,
and $Fermi/GBM$, we have applied the above methods to examine how
many classes may exist among them. We test the four kinds of
distribution for each class on the $logT_{90}-logHR$ plane, i.e.,
$\mathcal{N}$, $\mathcal{S N}$, $\mathcal{T}$, $\mathcal{S T}$.
The number of mixed components, i.e. classes, ranges from 1 to 5.
Table \ref{tab:table3} presents all the $BIC$ values, together
with the minimum value of $BIC_{min}$ which indicates the optimal
fitting model for the data set.

From Table \ref{tab:table3}, it can be seen that all three samples
support a symmetrical distribution ($\mathcal{N}$ or
$\mathcal{T}$) both for 1 time and $10^4$ times fitting, which
generally has a smaller $BIC$ as compared with skewed
distributions ($\mathcal{S N}$ or $\mathcal{S T}$). At the same
time, we could also see that the number of components should
generally be two or three, since the $BIC$ values are much higher
when the component number is taken as 1, 4, or 5.

For the $CGRO/BATSE$ sample, the preferred distribution is
$\mathcal{T}$, with three components, both for 1 time and $10^4$
times fitting. For $Swift/BAT$ GRBs, the optimal distribution is
$\mathcal{N}$, with three components, both for 1 time and $10^4$
times fitting. For $Fermi/GBM$ events, the optimal distribution is
$\mathcal{N}$ both for both 1 time and $10^4$ times fitting. Note
that the 1 time fitting prefers two components while $10^4$ times
fitting prefers three components.

However, it is worth mentioning that the difference in $BIC$ is
only $\Delta{BIC_{2\mathcal{N}-3\mathcal{N}}} = -1.968$ (5.317 for
$10^4$ times fitting) for 1 time fitting. When the number of free
parameters (as is shown in Table \ref{tab:table2}; indicating the
complexity of the model) of the fitting model ($3\mathcal{N}$ is
17, $2\mathcal{N}$ is 11) is considered, it is hard to say that
the three component classification scheme is better than the two
component classification scheme.

Additionally, from Table \ref{tab:table3}, we could see that the
second best fitting distribution is $\mathcal{N}$, $\mathcal{T}$,
$\mathcal{T}$ for $CGRO/BATSE$, $Swift/BAT$, $Fermi/GBM$ GRBs,
respectively, while the asymmetric distributions ($\mathcal{S N}$,
$\mathcal{S T}$) are generally unsupported.

\begin{figure}
    \includegraphics[width=\columnwidth]{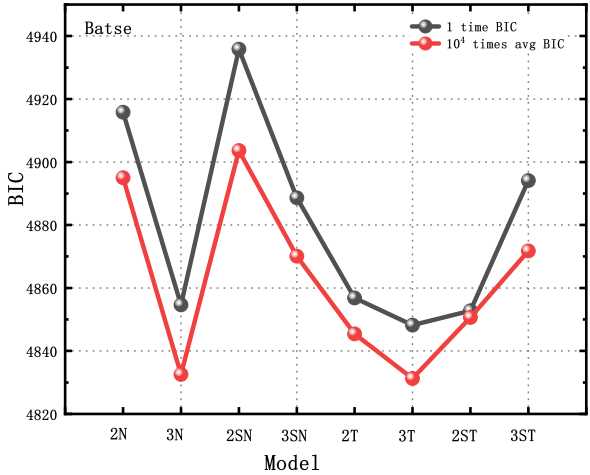}
    \caption{$BIC$ values of 1 time and $10^4$ times fitting results for the
             $CGRO/BATSE$ data set. The X-axis represents different models and
             the corresponding number of mixed components. The Y-axis represents
             $BIC$ values. The grey line corresponds to 1 time fitting, while
             the red line shows the average $BIC$ of $10^4$ times fitting results.
             }
    \label{fig:1_figure}
\end{figure}

Fig.\ref{fig:1_figure} shows the results of 1 time and $10^4$
times fitting for the $CGRO/BATSE$ data set. It can be seen that
the profiles of 1 time fitting and $10^4$ times fitting are quite
similar, but the $BIC$ of $10^4$ times fitting is generally
significantly lower. We could also see that for the 1 time
fitting, $3\mathcal{T}$ is the optimal fitting model, while
$3\mathcal{N}$ is the second optimal fitting model.  For the
$10^4$ times fitting,  3$\mathcal{N}$ becomes the optimal model
while 3$\mathcal{T}$ becomes the second optimal model. In all the
cases, the $\mathcal{S N}$ and $\mathcal{S T}$ distribution
(skewed models) can be safely expelled since their $BIC$ values
are obviously too high.

\begin{figure}
    \includegraphics[width=\columnwidth]{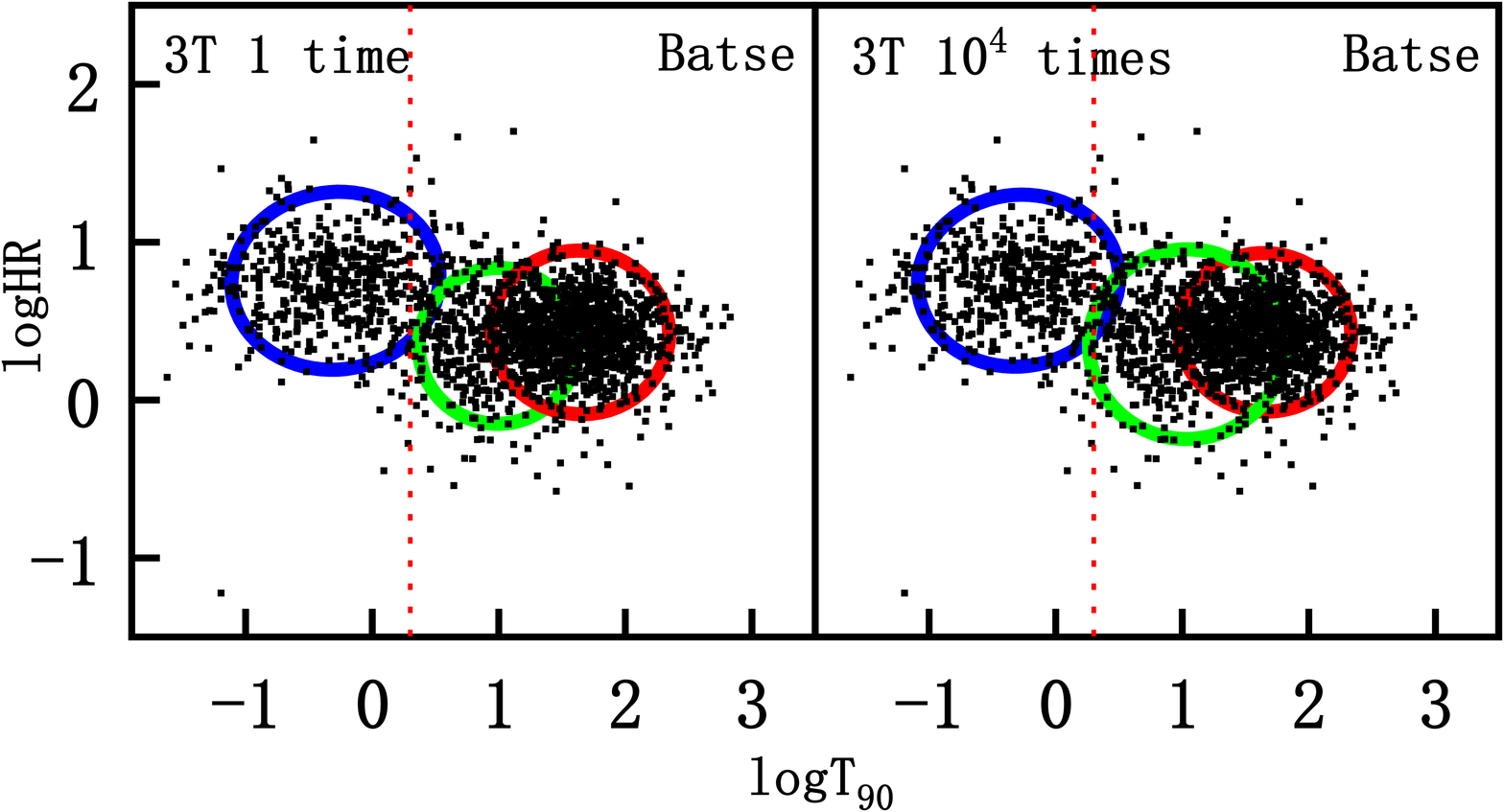}
    \caption{Fitting results of the $CGRO/BATSE$ data with the optimal model of $3\mathcal{T}$.
             Both the X-axis and Y-axis are in logarithmic coordinates. The left panel shows the 1 time
             fitting $FWHM$ of each component, and the right panel correspondingly shows the $10^4$ times
             fitting result. The red dashed line represents $T_{90}$ = 2 s. The contour of the middle
             component overlaps severely with the contour of long GRBs, but is well separated from short
             GRBs, with little overlap.
             }
    \label{fig:2_figure}
\end{figure}

Fig.\ref{fig:2_figure} illustrates the 1 time and $10^4$ times
fitting results by the optimal $3\mathcal{T}$ model for
$CGRO/BATSE$ GRBs. There are three contour curves in this figure,
which are the full width at half maximum (hereafter $FWHM$) for each
component. The blue contour includes traditional short GRBs, which
are located to the left of $T_{90}$ = 2 s line, with a relatively
large $HR$ value. The red contour contains most of the traditional
long GRBs with $T_{90}$ = 2 s and with a relatively smaller $HR$.
The green contour shows the intermediate component.
Note that most GRBs included in this component have a duration of
$T_{90}$ > 2 s, and their $HR$ value is also small. It means that
these GRBs are quite similar to traditional long GRBs. The third
row of Table \ref{tab:table4} presents the number of GRBs included
in each component for the left panel of Fig.\ref{fig:2_figure}. We
see that the number of GRBs in each component (short,
intermediate, long GRBs) is 500, 385 and 1069, respectively. It
should also be noted that the GRBs in the intermediate component
only account for a ratio of 26.479$\%$ of all long GRBs, as could
be seen from Table \ref{tab:table4}. So, the intermediate GRBs, if
exists, is only a small group.

\begin{figure}
    \includegraphics[width=\columnwidth]{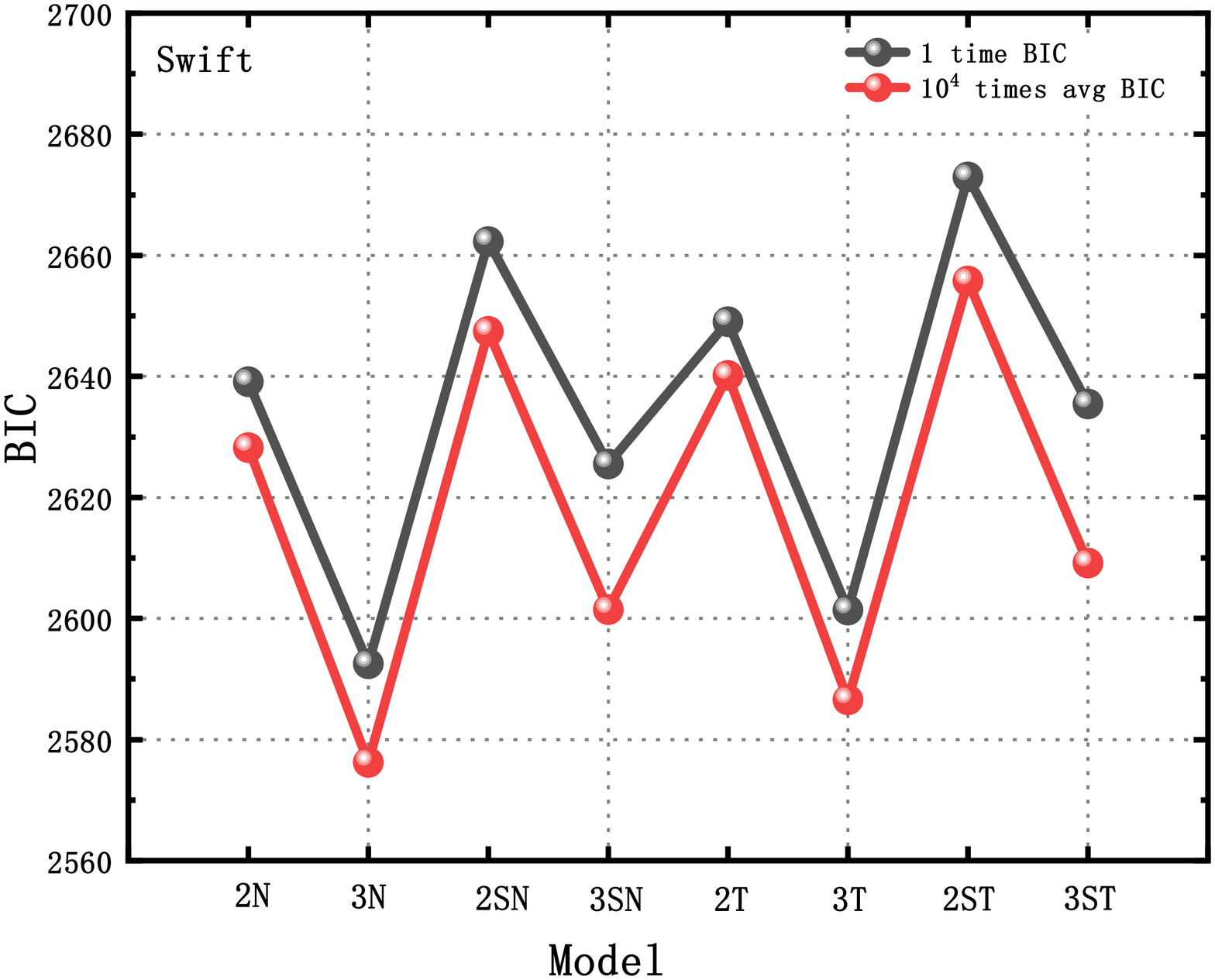}
    \caption{$BIC$ values of 1 time and $10^4$ times fitting results for
             the $Swift/BAT$ data set. The X-axis represents different models and
             the corresponding number of mixed components. The Y-axis represents
             $BIC$ values. The grey line corresponds to 1 time fitting, while
             the red line shows the average $BIC$ of $10^4$ times fitting results.
             Note that the 3$\mathcal{N}$ model gives the minimum $BIC$ value, thus
             presents the best fit.}
    \label{fig:3_figure}
\end{figure}

Fig.\ref{fig:3_figure} shows the 1 time and $10^4$ times fitting
results for the $Swift/BAT$ GRB data set. Similar to
Fig.\ref{fig:1_figure}, we see that the profiles of 1 time fitting
and $10^4$ times fitting are quite alike, with the $BIC$ of $10^4$
times fitting significantly lower. Fig.\ref{fig:3_figure} clearly
shows that $3\mathcal{N}$ is the optimal fitting model, while
$3\mathcal{T}$ is the second optimal fitting model. The
distribution of $\mathcal{S N}$ and $\mathcal{S T}$ can be
essentially expelled due to large $BIC$ values.

\begin{figure}
    \includegraphics[width=\columnwidth]{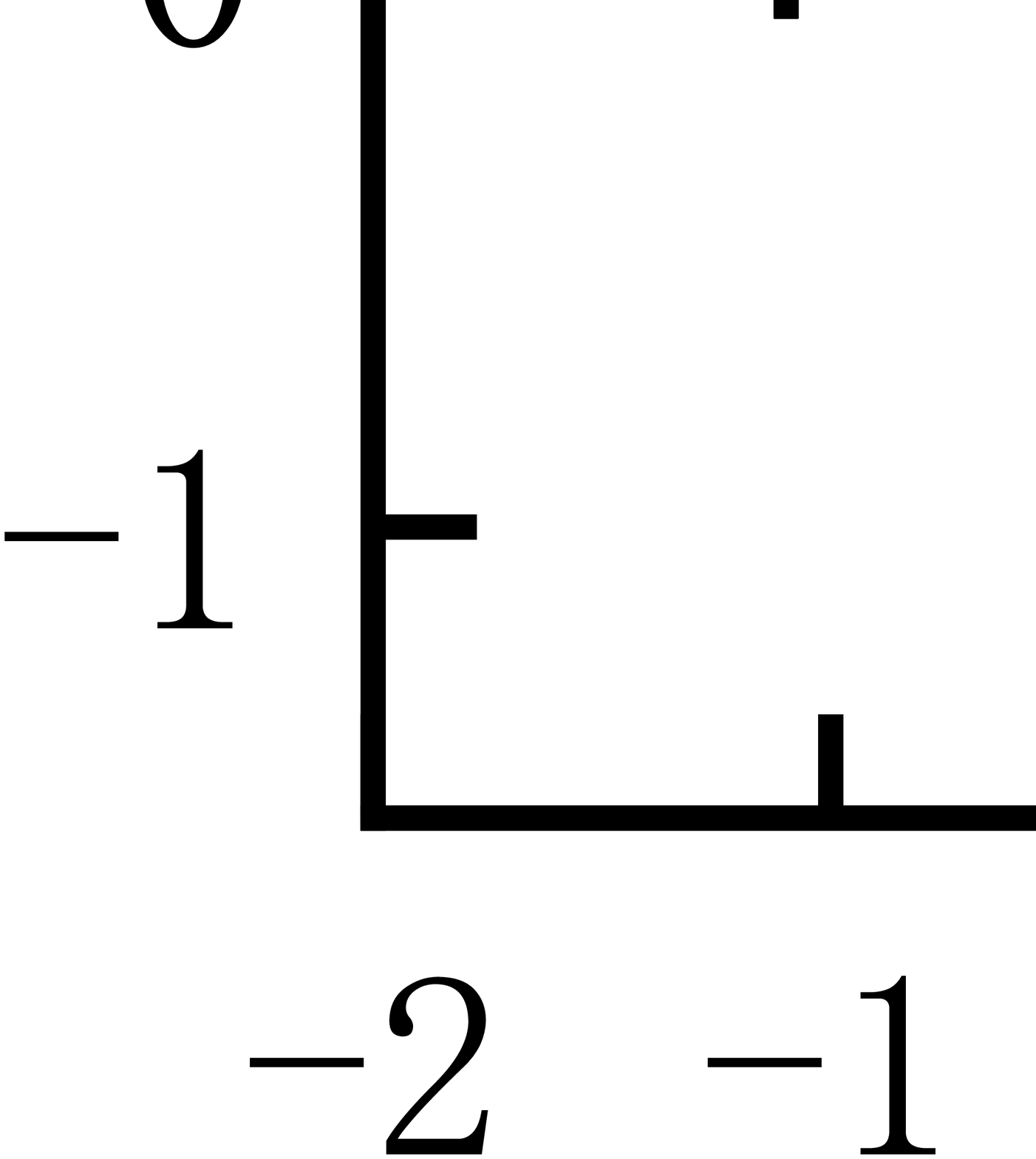}
    \caption{Fitting results of the $Swift/BAT$ data with the optimal model of $3\mathcal{N}$.
             Both the X-axis and Y-axis are in logarithmic coordinates. The left panel shows the 1 time
             fitting $FWHM$ of each component, and the right panel correspondingly shows the $10^4$ times
             fitting result. The red dashed line represents $T_{90}$ = 2 s. The contour of the middle
             component overlaps severely with the contour of long GRBs, but is well separated from short
             GRBs, with little overlap.
             }
    \label{fig:4_figure}
\end{figure}

Fig.\ref{fig:4_figure} illustrates the 1 time and $10^4$ times
fitting results by the optimal $3\mathcal{N}$ model for
$Swift/BAT$ GRBs. Again, there are three contour curves in
Fig.\ref{fig:4_figure}, which are the $FWHM$ for each component.
The blue contour includes traditional short GRBs. The red
contour contains most of the traditional long GRBs  The green contour
shows the intermediate component. Again, most GRBs of this component
are quite similar to traditional long GRBs, and the number of events
is small. From the fourth row of Table \ref{tab:table4},
it can be seen that the number of GRBs in each component (short,
intermediate, long GRBs) is 92, 507 and 766 respectively.
GRBs in the intermediate component only account for a ratio of
39.827$\%$ of all long GRBs, as shown in Table \ref{tab:table4}.
Note that this ratio is the largest among all the three data sets.

\begin{figure}
    \includegraphics[width=\columnwidth]{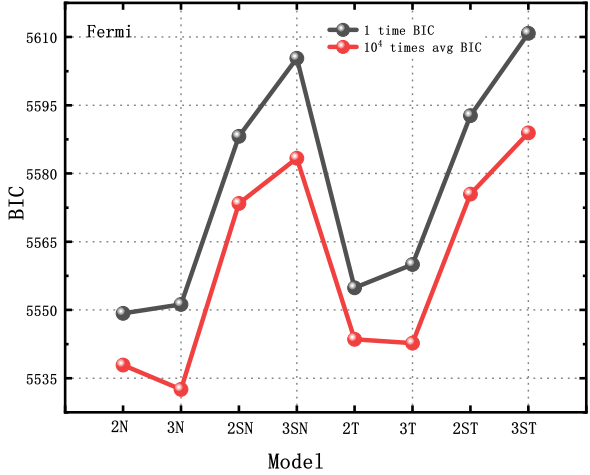}
    \caption{$BIC$ values of 1 time and $10^4$ times fitting results for
             the $Fermi/GBM$ data set. The X-axis represents different models and
             the corresponding number of mixed components. The Y-axis represents
             $BIC$ values. The grey line corresponds to 1 time fitting, while
             the red line shows the average $BIC$ of $10^4$ times fitting results.
             Note that the 2$\mathcal{N}$ model gives the minimum $BIC$ value for 1
             time fitting, while the 3$\mathcal{N}$ model gives the minimum $BIC$ for
             $10^4$ times fitting.}
    \label{fig:5_figure}
\end{figure}

Fig.\ref{fig:5_figure} shows the 1 time and $10^4$ times fitting
results for the $Fermi/GBM$ GRB data set. Again, the profiles of 1
time fitting and $10^4$ times fitting are quite similar, with the
$BIC$ of $10^4$ times fitting significantly lower. In
Fig.\ref{fig:5_figure}, we see that $3\mathcal{N}$ is the optimal
fitting model, and $3\mathcal{T}$ is the second optimal fitting
model. The distribution of $\mathcal{S N}$ and $\mathcal{S T}$ can
be safely ruled out since the corresponding $BIC$ values are
large.

\begin{figure}
    \includegraphics[width=\columnwidth]{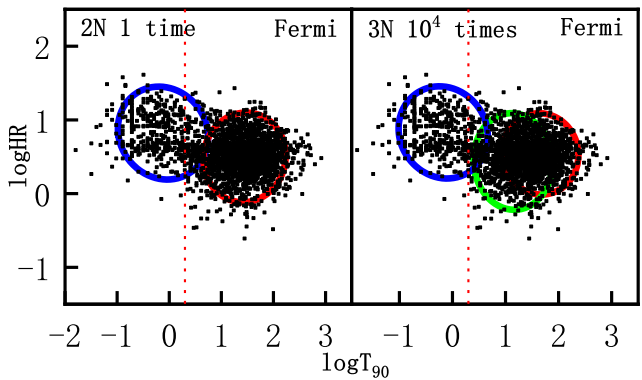}
    \caption{Fitting results of the $Fermi/GBM$ data with optimal model of $2\mathcal{N}$
             (for 1 time fitting) or $3\mathcal{N}$ (for $10^4$ times fitting).
             Both the X-axis and Y-axis are in logarithmic coordinates. The left panel shows the 1 time
             fitting $FWHM$ of each component, and the right panel correspondingly shows the $10^4$ times
             fitting result. The red dashed line represents $T_{90}$ = 2 s.
             In the right panel, the contour of the middle
             component overlaps severely with the contour of long GRBs, but is well separated from short
             GRBs, with little overlap.}
    \label{fig:6_figure}
\end{figure}

Fig.\ref{fig:6_figure} presents the results of optimal
$2\mathcal{N}$ model for 1 time fitting, and optimal
$3\mathcal{N}$ model for $10^4$ times fitting, for $Fermi/GBM$
GRBs. As a result, there are two contour curves in the left panel
and three contour curves in the right panel of
Fig.\ref{fig:6_figure}. The blue contour mainly includes short
GRBs. The red contour contains most of the long GRBs. The green
contour in the right panel corresponds to the intermediate
component, which are also long GRBs. Similar to $CGRO/BATSE$
events, the number of GRBs in this group is small. The first and
second row of Table \ref{tab:table4} shows the number of GRBs in
each component of Fig.\ref{fig:6_figure}. In the left panel of
Fig.\ref{fig:6_figure}, 461 and 1849 GRBs are included in the
short GRB group and long GRB group, respectively. In the right
panel, there are 419, 625, 1266 GRBs in the three groups,
respectively. Again, we see that GRBs in the intermediate
component only account for a ratio of 33.051$\%$ of all long GRBs,
as shown in Table \ref{tab:table4}.

\begin{table*}
\centering \caption{Number of GRBs in each component for all the
three samples, under the optimal distribution model.}
\label{tab:table4}
\begin{threeparttable}
\begin{tabular}{lcccccccc}
\hline
Detector & Total No. & GRBs with $T_{90}<2s$ & left No.\tnote{a} & middle No.\tnote{a} & right No.\tnote{a} & ratio1\tnote{b} & ratio2\tnote{b} & ratio3\tnote{b} \\
\hline
$Fermi/GBM (2\mathcal{N})$ & 2310 & 372 & 461 & - & 1849 & 123.925$\%$ & - & 100$\%$ \\
$Fermi/GBM (3\mathcal{N})$ & 2310 & 372 & 419 & 625 & 1266 & 112.634$\%$ & 33.051$\%$ & 66.949$\%$ \\
$CGRO/BATSE (3\mathcal{T})$ & 1954 & 466 & 500 & 385 & 1069 & 107.296$\%$ & 26.479$\%$ & 73.521$\%$ \\
$Swift/BAT (3\mathcal{N})$ & 1365 & 118 & 92 & 507 & 766 & 77.966$\%$ & 39.827$\%$ & 60.173$\%$ \\
\hline
\end{tabular}
\begin{tablenotes}
       \footnotesize
       \item[a] The left No. counts the number of GRBs included in the blue $FWHM$
                contour in the left panel of Fig.\ref{fig:2_figure}, Fig.\ref{fig:4_figure},
                Fig.\ref{fig:6_figure}; The middle No. counts that included in the
                green $FWHM$ contour; The right No. counts those included in the red $FWHM$ contour.
       \item[b] ratio1 = $\frac{left No.}{No. of GRBs with T_{90}<2s }$,
                ratio2 = $\frac{middle No.}{middle No. + right No.}$,
                ratio3 = $\frac{right No.}{middle No. + right No.}$.
     \end{tablenotes}
\end{threeparttable}
\end{table*}

\subsection{Weakness in the above direct analysis}
\label{sec:Comprehensive analysis of fitting results}

In the previous section, we have studied three GRB samples
observed by three detectors, examining the possibility that there
might be multiple groups (from 1 to 5 components) of GRBs.  For
each component, we assume that it could be one of the following
four distributions: $\mathcal{N}$, $\mathcal{S N}$, $\mathcal{T}$,
$\mathcal{S T}$. It is found that the optimal fitting results
prefer symmetric distributions ($\mathcal{T}$ or $\mathcal{N}$).
For $CGRO/BATSE$ and $Swift/BAT$ GRBs, the results support that
there are 3 components.  But for $Fermi/GBM$ GRBs, the difference
in $BIC$ for 2 components and 3 components is very small so that
both possibilities could exist.

However, it should be noted that there some complicated factors in
the three GRB samples, which might prevent us from drawing a firm
conclusion directly from the observational data:

(i) The three detectors work in different energy bands
\citep{2019ApJ...870..105T}, which may lead to different duration
of $T_{90}$ even for the same event. The derived $HR$ parameter
also may have different physical meanings.

(ii) The three detectors have different technical parameters,
such as the field of view, signal to noise ratio, sensitivity
\citep{2019ApJ...870..105T}. These may also lead to
significant systematic difference in the three samples.

(iii) For each detector, the sample size (i.e. the number of GRBs
contained in the data set) is also different from each other.
This may have an impact on the GRB classification analysis
\citep{1998ApJ...508..757H,2002A&A...392..791H,2015A&A...581A..29T,2019ApJ...870..105T}.

Both Factor (i) and Factor (ii) are connected to the selection
effect of instruments and observations. They can significantly
affect the classification of GRBs \citep{2013ApJ...763...15Q}. To
overcome this problem, we need to know the intrinsic wide band
spectra of GRBs as well as detailed response characteristics of
each instruments, which is beyond the scope of this study. Here we
will mainly concentrate on Factor (iii). The sample size may also
affect our conclusion on the classification.
\citet{1998ApJ...508..757H} investigated an early version of
$BATSE$ GRB catalogue, which contains 797 events. They found a
prominent third peak, between the short and long groups, in the
log $T_{90}$ distribution, and hence claimed the existence of an
intermediate-duration class of GRBs. Later, when an updated
$BATSE$ catalogue that contains 1929 events was re-analyzed, it
was found that the intermediate peak almost disappears
\citep{2002A&A...392..791H,2015A&A...581A..29T,2019ApJ...870..105T}.
\citet{2019ApJ...870..105T} argued that it may be due to the
skewness of the distribution of long GRBs.

We notice that the three sample sizes are 1365, 1954, 2310,
respectively. With the increasing of the sample size, the $BIC$
difference between two component fitting and three component
fitting becomes smaller, which are $\Delta{BIC_{2\mathcal{N(T)} -
3\mathcal{N(T)}}} = 46.584, 8.622, -1.968$ in the 1 time fitting
case, as shown in Table \ref{tab:table3}. It clearly indicates
that the classification scheme may be seriously affected by the
sample size: when the sample size is small, the intermediate
component may appear; but when the sample size becomes large, the
intermediate component seems to disappear and merge with long
GRBs. To overcome the problem induced by the limiting sample size
(Factor iii), we will use the Bootstrap resampling method
\citep{2016MNRAS.462.3243Z} to construct some meaningful large
samples, based on which we could further investigate the
classification of GRBs in detail.

\begin{table*}
\centering
\caption{$BIC$ values of the expanded samples generated
         through the bootstrap re-sampling method. The sample size has been
         expanded to 3225 GRBs for each detector. }
\label{tab:table5}
\begin{threeparttable}
\begin{tabular}{lcccccc}
\hline
\multirow{2}{*}{Detector} & \multicolumn{2}{c|}{Resample of 2T(N)} &  & \multicolumn{2}{c|}{Resample of 3T(N)} & \\
 & 2T(N) & 3T(N) & $\Delta{BIC} < 0$\tnote{a} & 2T(N) & 3T(N) & $\Delta{BIC} < 0$\tnote{a} \\
\hline
$CGRO/BATSE$ & 14560.829 & 14606.858 & 10000 & 14300.49 & 14340.361 & 9958 \\
$Fermi/GBM$ & 13692.152 & 13733.845 & 9990 &  13573.818 & 13611.664 & 9972 \\
$Swift/BAT$ & 12571.067 & 12615.573 & 9986 & 12603.524 & 12633.937 & 9608 \\
%$Swift/BAT (doubled)$\tnote{a} & 13349.147 & 13395.399 & 10000 & 13322.587 & 13344.864 & 9202 \\
%$Swift/BAT (tripled)$\tnote{b} & 14024.951 & 14068.803 & 10000 & 13767.812 & 13787.979 & 8846 \\
\hline
\end{tabular}
\begin{tablenotes}
       \footnotesize
       \item[] Note: For the $10^4$ times fitting, the $BIC$ here is the average value.
       \item[a] $\Delta{BIC} = BIC_{2\mathcal{N(T)}} - BIC_{3\mathcal{N(T)}}$.
                This column counts the number of fitting times that prefers the two components
                scheme over the three components scheme, after a total of $10^4$
                times fitting.
       %\item[b] $Swift/BAT (doubled)$ means to double the short $GRBs$ in $Swift/BAT$ samples ,while $Swift/BAT (tripled)$ means to triple the short $GRBs$ in $Swift/BAT$ samples. The meaning in the following is the same.
     \end{tablenotes}
\end{threeparttable}
\end{table*}

\begin{figure}
    \includegraphics[width=\columnwidth]{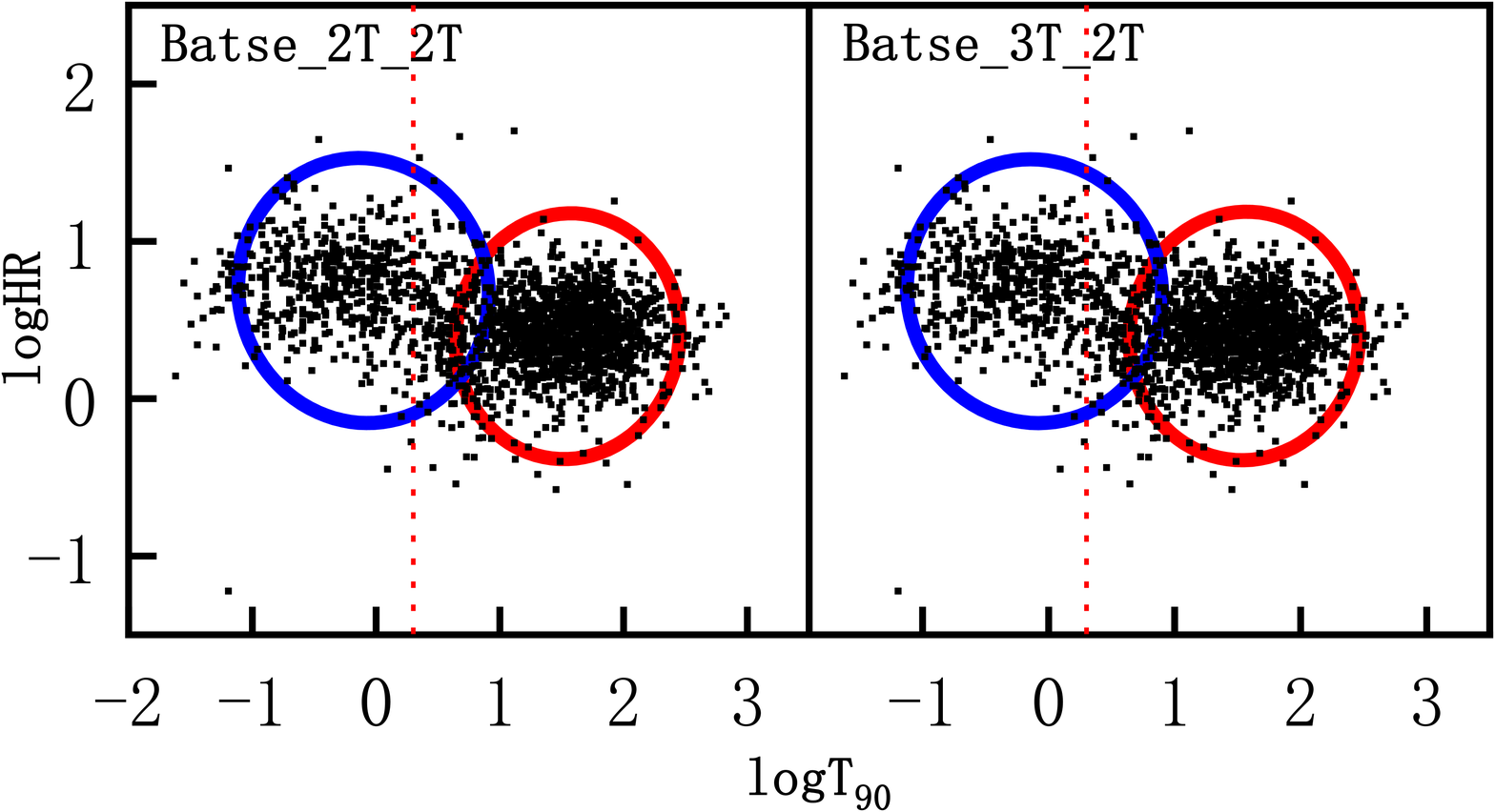}
    \caption{Illustration of our best fit to the mock $CGRO/BATSE$ sample
        (sample size now expanded to 3255 GRBs) with the $2\mathcal{T}$ model.
        Both the X-axis and Y-axis are in logarithmic coordinates.
        In the left panel, the 3255 mock GRBs are generated by
        assuming a $2\mathcal{T}$ distribution. In the right
        panel, the 3255 mock GRBs are generated by assuming a
        $3\mathcal{T}$ distribution. In both panels, the mock GRBs
        are finally best fit by the $2\mathcal{T}$ distribution after
        $10^4$ times bootstrap analysis, as shown by the blue and
        red $FWHM$ contours. In this figure, the black
        scatter points correspond to the real GRBs detected by
        $CGRO/BATSE$. The red dashed line represents
        $T_{90}$ = 2 s.
        }
    \label{fig:7_figure}
\end{figure}

\begin{figure}
    \includegraphics[width=\columnwidth]{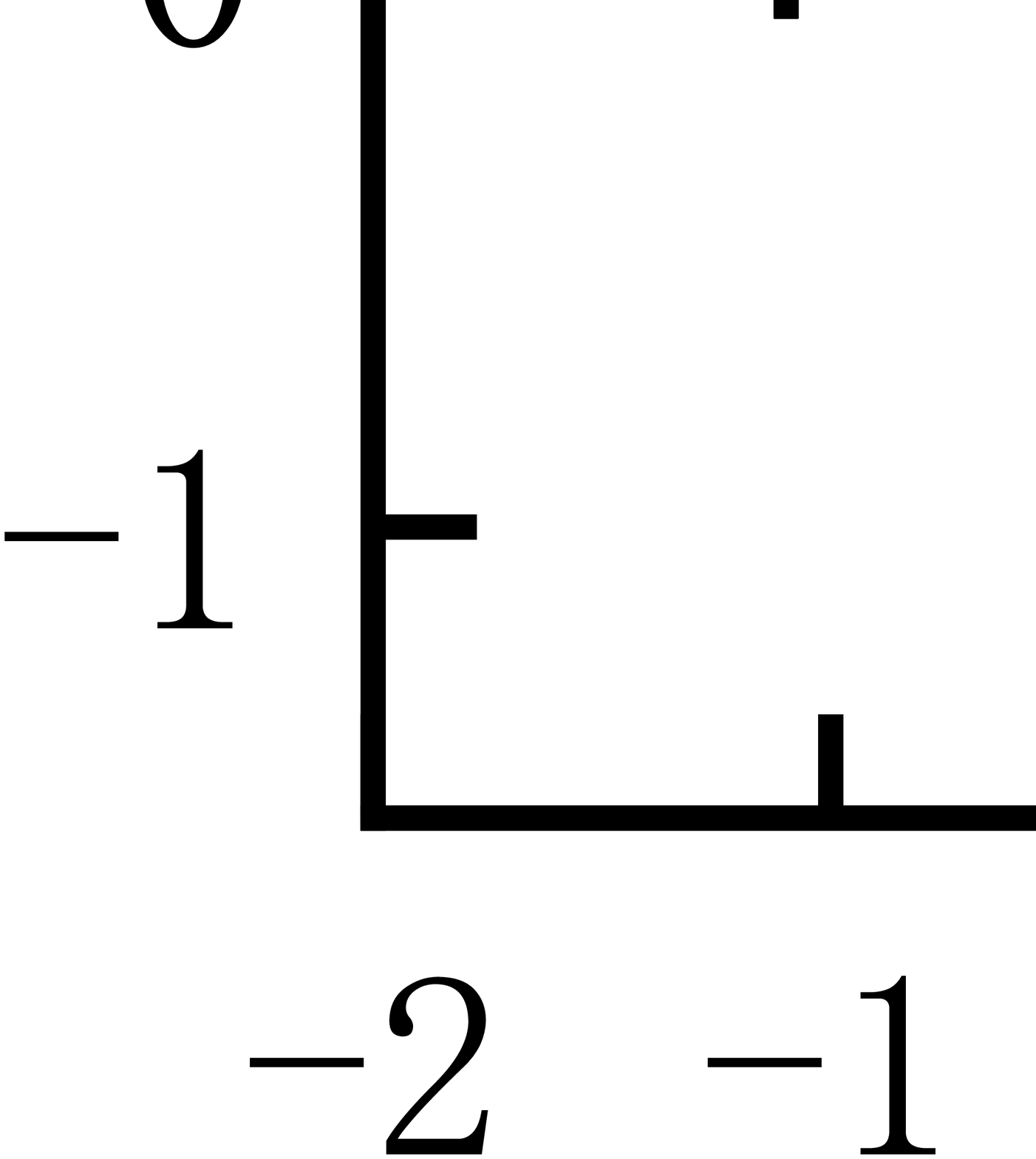}
    \caption{Illustration of our best fit to the mock $Swift/BAT$ sample
       (sample size now expanded to 3255 GRBs) with the $2\mathcal{N}$ model.
       Both the X-axis and Y-axis are in logarithmic coordinates.
       In the left panel, the 3255 mock GRBs are generated by
       assuming a $2\mathcal{N}$ distribution. In the right
       panel, the 3255 mock GRBs are generated by assuming a
       $3\mathcal{N}$ distribution. In both panels, the mock GRBs
       are finally best fit by the $2\mathcal{N}$ distribution after
       $10^4$ times bootstrap analysis, as shown by the blue and
       red $FWHM$ contours. In this figure, the black
       scatter points correspond to the real GRBs detected by
       $Swift/BAT$. The red dashed line represents
       $T_{90}$ = 2 s.
       }
    \label{fig:8_figure}
\end{figure}

\begin{figure}
    \includegraphics[width=\columnwidth]{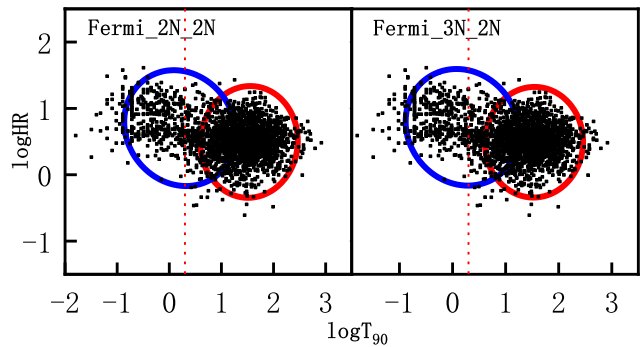}
    \caption{Illustration of our best fit to the mock $Fermi/GBM$ sample
       (sample size now expanded to 3255 GRBs) with the $2\mathcal{N}$ model.
       Both the X-axis and Y-axis are in logarithmic coordinates.
       In the left panel, the 3255 mock GRBs are generated by
       assuming a $2\mathcal{N}$ distribution. In the right
       panel, the 3255 mock GRBs are generated by assuming a
       $3\mathcal{N}$ distribution. In both panels, the mock GRBs
       are finally best fit by the $2\mathcal{N}$ distribution after
       $10^4$ times bootstrap analysis, as shown by the blue and
       red $FWHM$ contours. In this figure, the black
       scatter points correspond to the real GRBs detected by
       $Fermi/GBM$. The red dashed line represents
       $T_{90}$ = 2 s.
       }
    \label{fig:9_figure}
\end{figure}

\subsection{Bootstrap resampling analysis}
\label{sec:Bootstrap resampling analysis}

The bootstrap analysis is done in the following steps:

(i) Expand the number of GRBs detected by a particular detector to
3255 events. To do so, we use the optimal distribution function
($\mathcal{T}$ for $BATSE$; $\mathcal{N}$ for $Swift$ and $Fermi$)
and assume that the optimal classification scheme is either two
components or three components. Following the optimal distribution
function, we can randomly generate 3255 mock GRBs. This process is
usually referred to as re-sampling. Note that 3255 is simply the
number of GRBs detected by $Fermi/GBM$ as of March 28, 2022 (see
the HEASARC website), which is the largest sample size for GRBs
detected by the three detectors. In our simulations, we set 3255
as the destination sample size.

(ii) With the mock GRB sample, using two component model and three
component model to fit it to see which model presents a better
result. We use 2$\mathcal{T}$ and 3$\mathcal{T}$ to fit the mock
$BATSE$ sample, and use 2$\mathcal{N}$ and 3$\mathcal{N}$ to fit
the mock samples of $Swift$ and $Fermi$ detectors.

(iii) Repeat step (i) and (ii) for $10^4$ times, and count the
number of times that the mock samples are divided into two
components and three components separately.

The $BIC$ values of the mock samples are displayed in Table
\ref{tab:table5}. The best fit results are illustrated in
Fig.\ref{fig:7_figure} --- Fig.\ref{fig:9_figure}. From Table
\ref{tab:table5} and the figures, we find that for the mock
$BATSE$ samples generated from 2$\mathcal{T}$ model, 100$\%$
of them are again better fit by the 2$\mathcal{T}$ distribution.
For the mock $Fermi$ samples generated from 2$\mathcal{N}$ model,
99.9$\%$ of them are again better fit by the 2$\mathcal{N}$
distribution. For the mock $Swift$ samples generated from
2$\mathcal{N}$ model, 99.86$\%$ of them are again better
fit by the 2$\mathcal{N}$ distribution.

On the other hand, for the mock $BATSE$ samples generated from
3$\mathcal{T}$ model, 99.58$\%$ of them are better fit by the
2$\mathcal{T}$ distribution. For the mock $Fermi$ samples
generated from 3$\mathcal{N}$ model, 99.72$\%$ of them are better
fit by the 2$\mathcal{N}$ distribution. For the mock $Swift$
samples generated from 3$\mathcal{N}$ model, 96.08$\%$ of them are
again better fit by the 2$\mathcal{N}$ distribution. We see that
after expanding the sample to 3225 GRBs, even though the mock GRBs
are originally assumed to follow 3-component distribution, the
sample finally is still better described by a 2-component
classification. It strongly indicates that a large sample size
tend to support a two-component scheme.

\begin{table}
\centering \caption{Ratio of short GRBs in each sample.}
\label{tab:table6}
\begin{tabular}{lccc}
\hline
Detector & total No. & No. of short GRBs & Ratio of short GRBs \\
\hline
$CGRO/BATSE$ & 1954 & 466 & 23.8$\%$ \\
$Fermi/GBM$ & 2310 & 372 & 16.1$\%$ \\
$Swift/BAT$ & 1365 & 118 & 8.6$\%$ \\
%$Swift/BAT (doubled)$ & 1484 & 236 & 15.9$\%$ \\
%$Swift/BAT (tripled)$ & 1602 & 354 & 22.1$\%$ \\
\hline
\end{tabular}
\end{table}

As shown in Fig.\ref{fig:7_figure} and Fig.\ref{fig:9_figure},
the $FWHM$ contours derived from the bootstrap analysis can
well identify the two components in the sample on the
$logT_{90}-logHR$ plane for mock GRBs of $BATSE$ and $Fermi$
detectors. However, in Fig.\ref{fig:8_figure}, the blue contour
seems to obviously deviate from the short component. As a result,
many short $Swift$ bursts are outside the contour circle. This
may be due to the fact that the number of short GRBs is relatively
small in the $Swift$ sample, resulting in a low fitting weight.
Table \ref{tab:table6} lists the ratios of short GRBs in each
sample. For $BATSE$ GRBs, the number of short bursts is 466,
accounting for 23.8$\%$ of the total events. For the $Fermi$
sample, there are 372 short bursts and the ratio is 16.1$\%$.
But in the $Swift$ sample, these are only 118 short events,
which corresponds to a small ratio of 8.6$\%$.

\section{CONCLUSION}

\label{sec:CONCLUSION}

In this study, we investigate the distribution and classification
of $BATSE$, $Swift$ and $Fermi$ GRBs. The possible existence of up
to 5 components is carefully examined by using the Bayesian
information criterion on the $logT_{90}$ -- $logHR$ plane. For each
component, the distribution could be one of the four forms:
$\mathcal{N}$, $\mathcal{S N}$, $\mathcal{T}$, or $\mathcal{S
T}$. Generally, all the three samples show a symmetric (either
$\mathcal{N}$ or $\mathcal{T}$) distribution. For $CGRO/BATSE$ and
$Swift/BAT$ data sets, the best fitting models are $3\mathcal{T}$
and $3\mathcal{N}$, respectively. However, for the $Fermi/GBM$
sample, $2\mathcal{N}$ is almost as good as $3\mathcal{N}$. It is
also found that the sample size has a great impact on the
classification. When the sample size is large, the best
classification scheme universally tends to two, instead of three
as in the small sample size cases. A further bootstrap re-sampling
analysis strongly supports this result.

\citet{2019MmSAI..90...45T,2019ApJ...870..105T} have used a
mixture of 2 --- 3 components of $\mathcal{N}$, $\mathcal{S N}$,
$\mathcal{T}$, $\mathcal{S T}$ to fit the data sets of
$CGRO/BATSE$, $Fermi/GBM$, $Swift/BAT$, $Konus-Wind$, $RHESSI$,
$Suzaku/WAM$ on the $logT_{90}$ -- $logHR$ plane. They argued that
the 2$\mathcal{S T}$ model is better in describing the
$CGRO/BATSE$ and $Fermi/GBM$ data sets under both $AIC$ and $BIC$
criteria \citep{2019ApJ...870..105T}. For $Swift$, they claimed
that the $BIC$ criteria prefers 2$\mathcal{S T}$ or 2$\mathcal{S
N}$, while the $AIC$ prefers 3$\mathcal{S T}$ or 3$\mathcal{S N}$
\citep{2019MmSAI..90...45T}.

For the $CGRO/BATSE$ detector, we have adopted the same data set
as \citet{2019ApJ...870..105T}, but drawn a different conclusion
on the skewness of the distribution (our $10^4$ times fitting
results prefer 2$\mathcal{T}$, as compared with their 2$\mathcal{S
T}$). We notice that \citet{2019ApJ...870..105T} carried out their
studies mainly based on 1 time fitting, which may lead to
significant random fluctuations. In fact, from Table
\ref{tab:table3}, we could find that our 1 time fitting result of
$CGRO/BATSE$ is somewhat similar to that of
\citet{2019ApJ...870..105T}: 2$\mathcal{S T}$ is also a good fit,
which is only slight worse than the best fitting model of
3$\mathcal{T}$. In this study, we have further conducted $10^4$
times fitting analysis, which can effectively overcome the
fluctuations. We see that after $10^4$ bootstrap re-sampling and
fitting, the best fitting model is 2$\mathcal{T}$.

As for $Fermi/GBM$ and $Swift/BAT$ data sets, the difference
between our results and that of
\citet{2019MmSAI..90...45T,2019ApJ...870..105T} may be caused by
several factors. First, the sample size is different, and our
sample sizes are larger. The $Fermi/GBM$ data set contains merely
1376 GRBs in \citet{2019ApJ...870..105T}, but it is 2310 in this
study. Similarly, the $Swift/BAT$ data set contains 1033 events in
\citet{2019MmSAI..90...45T}, and it is 1365 in our work. Second,
the $HR$ parameter is calculated by different method. In our
study, this parameter is derived by using the optimal spectral
fitting model. Thirdly, as mentioned above, we use the $10^4$
times fitting analysis to overcome the fluctuations caused by 1
time fitting method. Our results on these two data sets further
strengthen the conclusion that a symmetric distribution
($\mathcal{N}$ or $\mathcal{T}$) is better than an asymmetric
distribution ($\mathcal{S N}$ or $\mathcal{S T}$) for the
currently observed GRBs on the $T_{90} - HR$ plane.

A symmetric lognormal distribution ($\mathcal{N}$ and $\mathcal{T}$)
may originate from realistic processes involving a number of
independent parameters which somewhat randomly distributed
in particular ranges \cite{1957Paper,1982Paper}. Many examples
of lognormal distributions are observed in nature, such as the
propagation of a laser beam in a turbulent
atmosphere \citep{1982Statistical}, the size of cumulus clouds
in the atmosphere, the strength of terrestrial lightning, etc.
Interestingly, for the soft gamma-ray repeating source of
$SGR$ 1806-20, the time intervals between bursts also follow
a lognormal distribution \citep{1994A&A...288L..49H}.
%And statistics lognormal distribution has been investigated by  the high-quality data of $BATSE$, $Swift$, $Fermi$.
Similarly, the symmetric lognormal distribution of GRBs may
be interpreted in the framework of the collapsar scenario:
the GRB duration may be determined by the ejected envelope
mass which itself could follow a symmetric
distribution \citep{2015Ap&SS.357....7Z}. Accretion of the
ejected envelope by the newly formed black hole may give
birth to the GRB. In the future, more GRBs will be observed
and the classification of them will be investigated in further
detail.

\section*{Acknowledgements}

We thank the anonymous referee for helpful comments and suggestions.
This study is supported by the National Natural Science Foundation
of China (Grant Nos. 12233002, 11873030, 12041306, 12147103, U1938201),
by the Youth Science \& Technology Talents Development Project of
Guizhou Education Department (No.KY[2022]098), by National SKA
Program of China No. 2020SKA0120300, by the National
Key R\&D Program of China (2021YFA0718500), and by the science
research grants from the China Manned Space Project with NO. 
CMS-CSST-2021-B11.

\section*{Data availability}

No new data were generated or analysed in support of this research.



\bibliographystyle{mnras}
\bibliography{Distribution_of_Gamma-Ray_Bursts-HYFrev2} % if your bibtex file is called example.bib

%\begin{thebibliography}{99}
%\end{thebibliography}

%%%%%%%%%%%%%%%%%%%%%%%%%%%%%%%%%%%%%%%%%%%%%%%%%%

%%%%%%%%%%%%%%%%% APPENDICES %%%%%%%%%%%%%%%%%%%%%

% Don't change these lines
\bsp    % typesetting comment
\label{lastpage}
\end{document}